\documentstyle[pra,multicol,aps,amstex,subeqnar,psfig]{revtex} 

\begin{document}

\newlength{\figwidtha}
\newlength{\figwidthb}
\setlength{\figwidtha}{0.5\linewidth}
\setlength{\figwidthb}{\linewidth}

\setlength{\textwidth}{150mm}
\setlength{\textheight}{240mm}
\setlength{\parskip}{2mm}

\input{epsf.tex}
\epsfverbosetrue


\renewcommand{\baselinestretch}{1.0}

\title{Escape angles in bulk $\chi^{(2)}$ soliton interactions}

\author{Steffen Kj\ae r Johansen, Ole Bang and Mads Peter S\o rensen}

\address{Department of Mathematical Modelling, Technical University 
of Denmark, DK-2800 Lyngby, Denmark}

\maketitle

\normalsize

\begin{abstract}
We develop a theory for non-planar interaction between two identical type 
I spatial solitons propagating at opposite, but {\em arbitrary transverse 
angles} in quadratic nonlinear (or so-called $\chi^{(2)}$) bulk media.
We predict quantitatively the outwards escape angle, below which the 
solitons turn around and collide, and above which they continue to move 
away from each other.
For in-plane interaction the theory allows prediction of the {\em outcome of 
a collision} through the inwards escape angle, i.e. whether the solitons 
fuse or cross. We find an analytical expression
determining the inwards escape angle using Gaussian 
approximations for the solitons. The theory is verified numerically.
\end{abstract}


\begin{multicols}{2}

Stable self-guided laser beams or optical bright spatial solitons are of 
substantial interest in basic physics\cite{Seg98} and for technical 
applications, such as inducing fixed\cite{Klo99} and dynamically 
reconfigurable waveguides\cite{Fue91Shi97Dit99}. 
Several types of spatial solitons have been demonstrated experimentally, 
including 1D Kerr solitons\cite{Ait90} and 1D and 2D (number of transverse 
dimensions) solitons in saturable\cite{Tik95},
photorefractive\cite{Dur93Dur95}, and $\chi^{(2)}$
media\cite{Sch96Tor95}.
Even incoherent solitons, excitable by a light bulb, have been
demonstrated in photorefractive media\cite{Mit96Mit97}.
All solitons exist when diffraction is balanced by the nonlinear 
self-focusing effect. In bulk Kerr media the self-focusing effect dominates
and leads to collapse of both coherent and incoherent 2D
solitons\cite{Her64Ban99}, their existence requiring an effectively
saturable nonlinearity.

\begin{figure}
  \centerline{\hbox{
  \psfig{figure=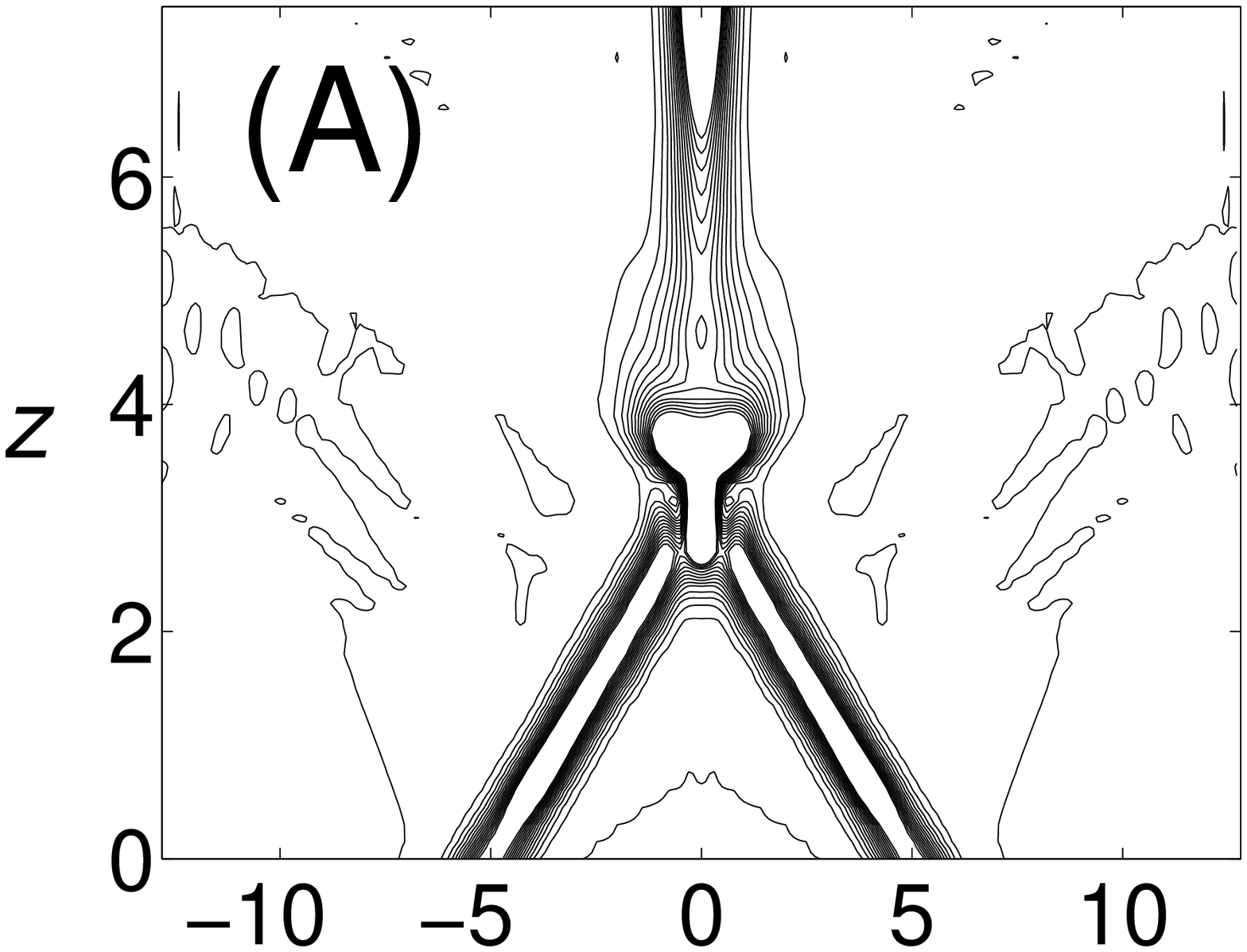,width=3.5cm,height=3.0cm}
  \psfig{figure=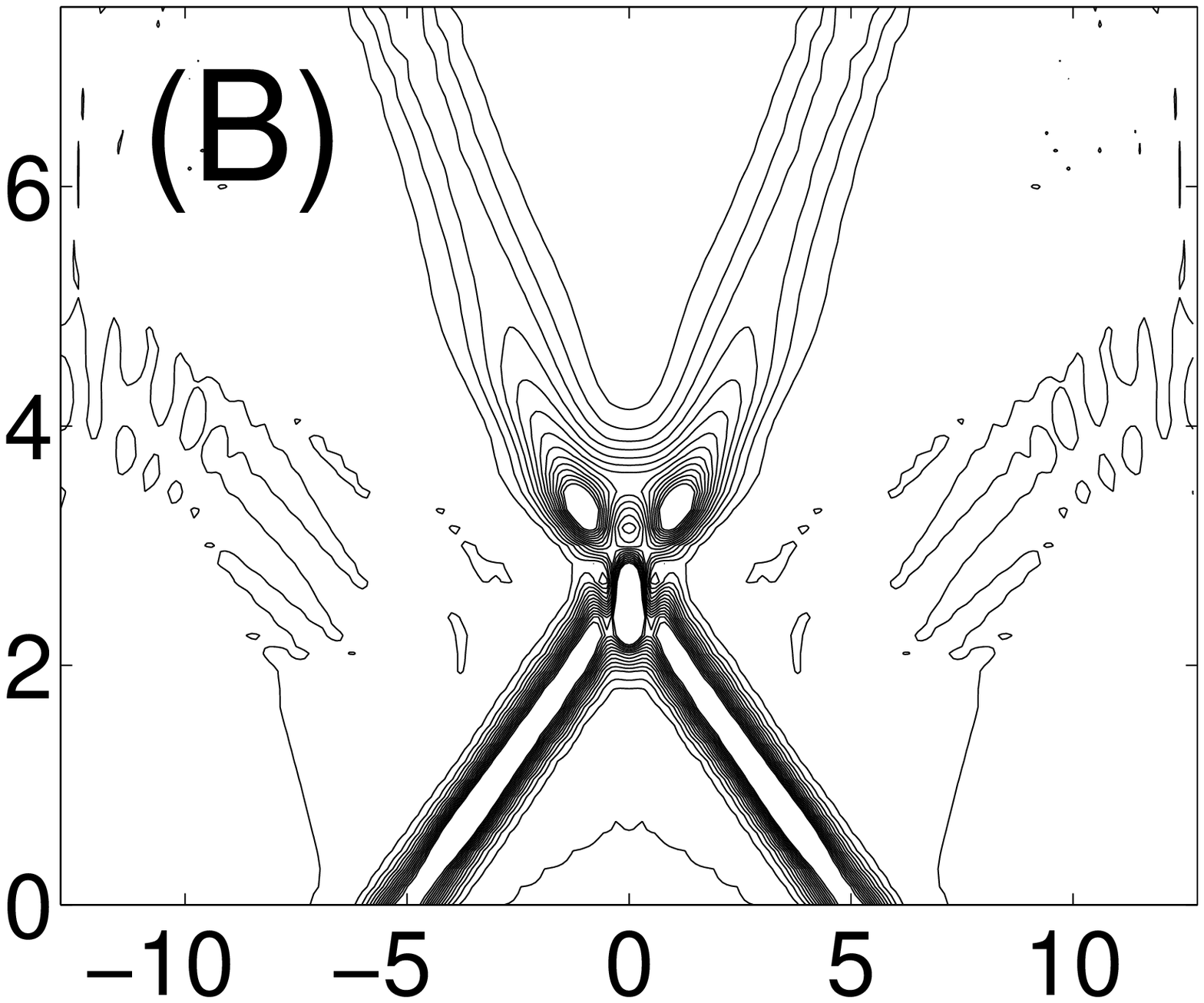,width=3.5cm,height=3.0cm}}}
  \centerline{\hbox{
  \psfig{figure=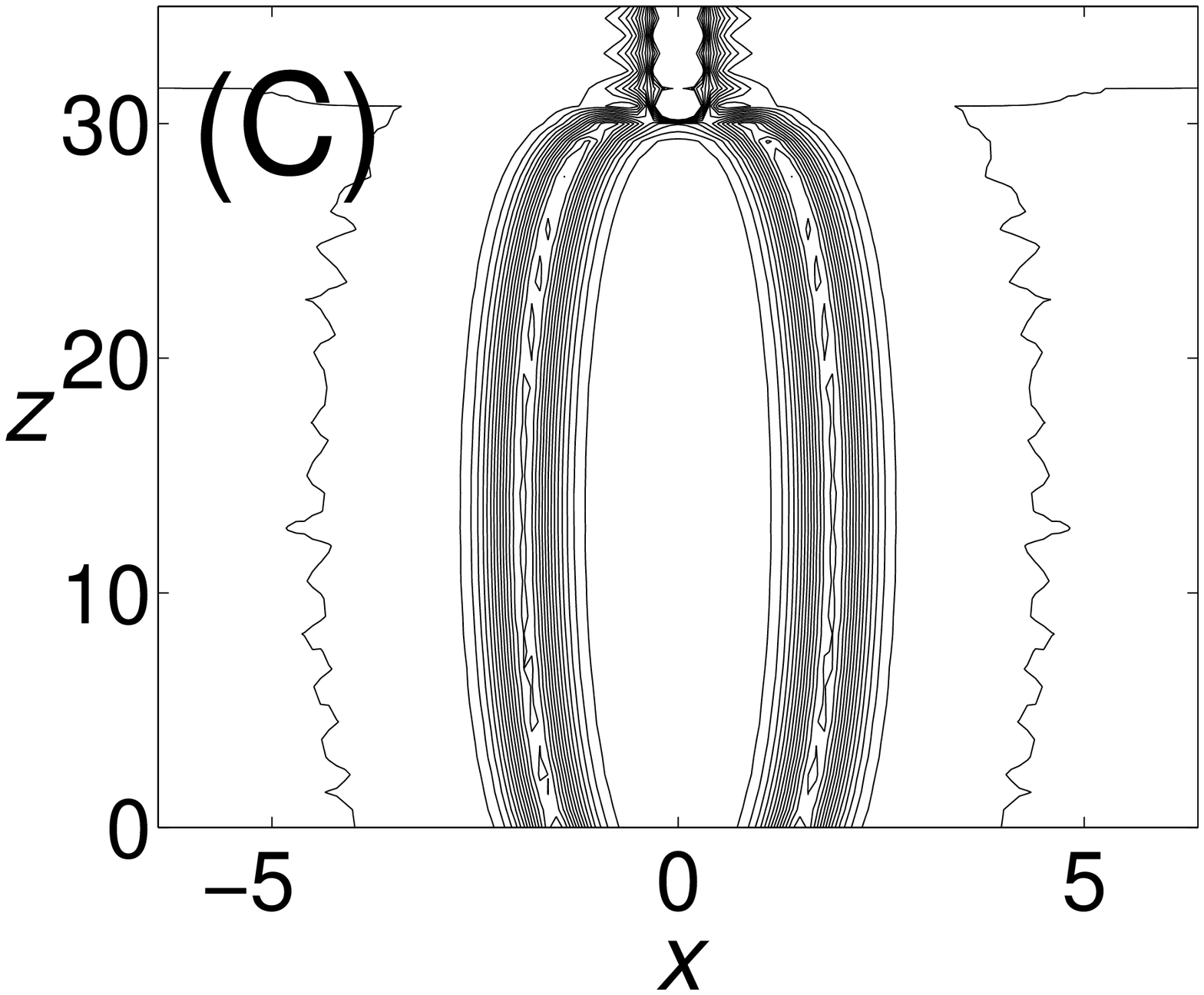,width=35mm,height=30mm}
  \psfig{figure=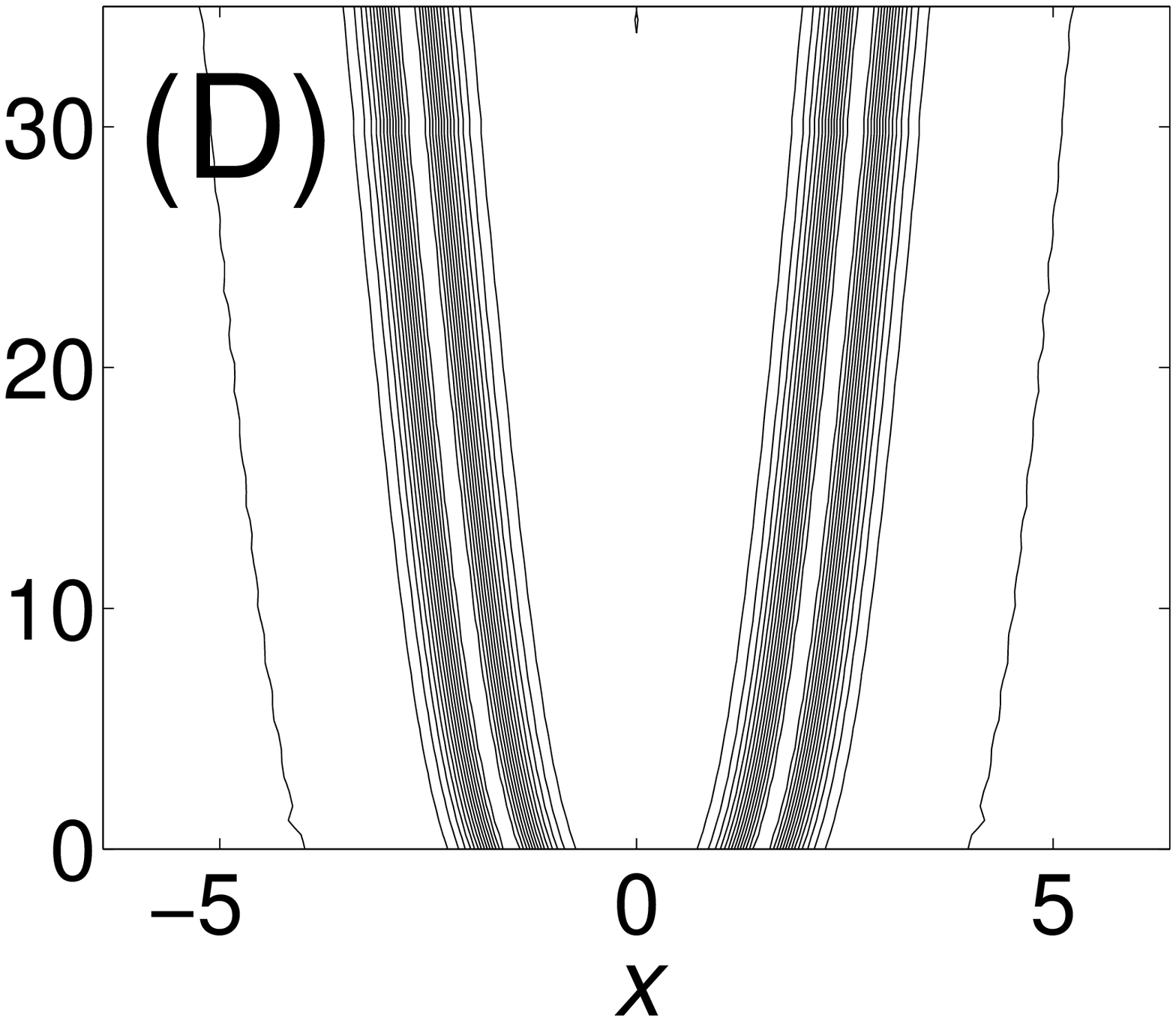,width=35mm,height=30mm}}\vspace{2mm}}
  \caption{A,B: Planar collision between two Gaussians, $P=48.3$ and
  $\beta=0$. C,D: Outwards launched exact solitons, $P=122.4$ and $\beta=5$. 
  The launch angles are $\alpha_x=58^\circ$ (A), $\alpha_x=62^\circ$ (B),  
  $\alpha_x=5.4^\circ$ (C), and $\alpha_x=5.7^\circ$ (D).}
  \label{fig:gauss_beta0}
\end{figure}

An intriguing feature of solitons is their particle like behavior during 
collision. In 1D Kerr media collisions are fully elastic due to integrability 
of the 1D nonlinear Schr\"odinger (NLS) equation\cite{Chi64Zak72}. 
In contrast, saturable, photorefractive, and $\chi^{(2)}$ media are 
described by non-integrable equations and soliton collisions are therefore 
inelastic, displaying both fusion (Fig. \ref{fig:gauss_beta0}:A), crossing
(Fig. \ref{fig:gauss_beta0}:B), repulsion, and annihilation, additional 
solitons can be generated in a fission-type process \cite{Kro97}, and 
solitons can even spiral around each other \cite{Shi97}. 
All processes depend strongly on the relative phase and have been 
demonstrated experimentally (see \cite{Seg98} and references therein).

Complex waveguide structures can be generated by soliton interaction, such
as directional couplers\cite{Lan99a}, but their efficient implementation
requires a detailed understanding of the nature of soliton collisions.
Snyder and Sheppard predicted the outcome of collisions of 1D solitons in 
saturable media by comparing the collision angle with the critical angle 
for total internal reflection in an equivalent waveguide\cite{Sny91}.
Except for this work most theories are based on the variational 
approach, which require the solitons to be far apart and breaks down
at collision.

Here we focus on $\chi^{(2)}$ materials \cite{Tor98}, which are more
general than the simpler cubic Kerr and saturable media in the sense 
that dependent on the phase-mismatch between the fundamental and 
second-harmonic (SH) waves the nonlinearity can be both purely 
quadratic (close to phase-matching) and effectively cubic (for a
large phase-mismatch).
Spatial solitons in $\chi^{(2)}$ materials do not modify the refractive 
index, and  consist of one (type I) or two (type II) fundamental fields 
resonantly coupled to a SH.
The $\chi^{(2)}$ materials are of significant interest to photonics due to the 
strong and fast nonlinearities they can provide through cascading\cite{Ste96}.
Furthermore, soliton induced waveguides in photorefractives can have a strong
$\chi^{(2)}$ nonlinearity, which can be used for second-harmonic generation
(SHG)\cite{Lan99b}.
Fusion and crossing (Fig. \ref{fig:gauss_beta0}:A,B) of spatial $\chi^{(2)}$ 
solitons has been demonstrated numerically\cite{Wer93Leo97} and experimentally 
\cite{Bae97Con98}, and fission of 1D type I solitons was demonstrated 
analytically and numerically in the large phase-mismatch limit approximately 
described by the NLS equation\cite{Tor96}.
However, the $\chi^{(2)}$ system is more general and complex than the saturable
NLS equation and so far variational theories were only able to predict critical
launch angles and relative phases separating regimes of collision and no 
collision \cite{Cla97Con97}. 
Elegant non-planar effective particle theories predicted the absence of 
spiraling type I solitons, but still required weakly overlapping solitons 
\cite{Ste98Bur99}.
In this letter we extend the effective particle approach to {\em arbitrary 
launch angles} and present the first theory able to correctly predict
the {\em outcome of collisions} between 2D type I solitons in $\chi^{(2)}$ 
media.

We consider beam propagation under type-I SHG conditions in lossless bulk 
$\chi^{(2)}$ materials. Neglecting walk-off the system of normalized
dynamical equations for the slowly varying envelope of the fundamental 
wave, $E_1=E_1(\vec{r}\,)$, and its SH, $E_2=E_2(\vec{r}\,)$, are 
\cite{Men94Ban97}
\begin{subeqnarray}\label{sys:chi2system}
&& i \partial_zE_{1}^{}
  +{\textstyle \frac{1}{2}}\nabla^2_\perp E_1
  +E_1^{*}E_2^{}=0,
\slabel{eqn:chi23system_a}\\
&& i\partial_zE_{2}^{}
  +{\textstyle \frac{1}{4}}\nabla^2_\perp E_2^{}
  -\beta E_2^{}
  +E_1^{2}=0.\slabel{eqn:chi23system_b}
\end{subeqnarray}
Here $\vec{r}$=$(x,y,z)$, $z$ is the propagation variable, and 
$\nabla^2_\perp=\partial_x^2+\partial_y^2$ accounts for diffraction 
in the transverse $\vec{r}_\perp=(x,y)$-plane. 
The normalized phase mismatch is  $\beta=l_d(2k_1-k_2)$, where $l_d$ 
is the diffraction length of the fundamental and $k_1$ and $k_2$ are the
wave numbers of the fundamental and SH, respectively. 
The system (\ref{sys:chi2system}) can be derived from
the Lagrangian density
\begin{eqnarray}
\cal{L}&=& 2\mbox{Im}\left(E_1\partial_zE^*_{1}\right)
        +\mbox{Im}\left(E_2\partial_zE^*_{2}\right)
        +\beta|E_2|^2\nonumber\\
      &&+|\nabla_\bot E_1|^2
        +{\textstyle \frac{1}{4}}|\nabla_\bot E_2|^2_{}
        -\mbox{Re}\left(E_2^*E_1^2 \right),
\label{eqn:lagrangiandens}
\end{eqnarray}
and conserves the power $P$=$\int(|E_1|^2$$+$$|E_2|^2)d\vec{r}_\perp$ 
and momentum $\vec{M}$=$\int\mbox{Im}\{E_1^*\nabla_\perp E_1+ 
\frac{1}{2}E_2^*\nabla_\perp E_2\}d\vec{r}_\perp$, where we have 
defined $\int d\vec{r}_\perp\equiv\int\int_{-\infty}^{\infty}dxdy$.

The system (\ref{sys:chi2system}) is known \cite{Bur95,Ste95} to have a 
\emph{one-parameter family} of radially symmetric bright 2D solitons of 
the form $E_1(\vec{r}\,)=V(r;\lambda)\exp(i\lambda z)$ and
$E_2(\vec{r}\,)=W(r;\lambda)\exp(i2\lambda z)$ where 
$\lambda>\mbox{max}(0;-\beta/2)$ is the internal soliton parameter and 
$r=\sqrt{x^2+y^2}$. 
We have found this family numerically, using a standard relaxation 
method, and approximately, using the variational approach \cite{Ste95} 
with Gaussians profiles $(V,W)=(V_g,W_g)$,
\begin{eqnarray}
  V_g = a_1 \exp(-r^2/b), \;
  W_g = a_2 \exp(-r^2/b).
  \label{eqn:gaussapp}
\end{eqnarray}
Here $a_1=a_2[2(\lambda b-1)]^{-\frac{1}{2}}$,
$a_2=\frac{3}{2}(\lambda+b^{-1})$, and $b=[1+(12 \lambda^2 +
8\lambda\beta+\beta^2)^{\frac{1}{2}}/(2\lambda+\beta)]/2\lambda$.
Because the system is Galilean invariant we can apply a gauge
transformation to find moving solitons. Thus the \emph{general
three parameter soliton family} $(\tilde{V},\tilde{W})$ is given by
\begin{subeqnarray}\label{sys:movingsolform}
  &&\tilde{V}(x-\nu_x z,y-\nu_y z;\lambda,\nu_x,\nu_y)=\nonumber\\ 
  &&\quad\quad V(r;\lambda-{\textstyle \frac{1}{2}}\nu_x^2-
  {\textstyle \frac{1}{2}}\nu_y^2)\exp[-i(\nu_xx+\nu_yy)]
  ,\qquad\slabel{eqn:newsolform1}\\ 
  &&\tilde{W}(x-\nu_x z,y-\nu_y z;\lambda,\nu_x,\nu_y)=\nonumber\\ 
  &&\quad\quad W(r;\lambda-{\textstyle \frac{1}{2}}\nu_x^2-
  {\textstyle \frac{1}{2}}\nu_y^2)\exp[-2i(\nu_xx+\nu_yy)]
  ,\qquad\slabel{eqn:newsolform2}
\end{subeqnarray}
where $(V,W)$ are either the exact soliton profiles
$(V_s,W_s)$  found numerically or $(V_g,W_g)$ given
by (\ref{eqn:gaussapp}). $\nu_{x,y}=\tan(\alpha_{x,y})$ are the initial
transverse velocities corresponding to the launch angles
$\alpha_{x,y}$ with respect to the $z$-axis.

We substitute a field composed of two weakly overlapping solitons
$(\tilde{V}^{(i)},\tilde{W}^{(i)})$ into the Lagrangian density 
(\ref{eqn:lagrangiandens}). 
We then follow the procedure outlined in \cite{Ste98Bur99} and
allow the solitons to vary adiabatically through a slow variation of the
soliton parameters with $Z=\epsilon z$ being the slow propagation variable. 
To first order in $\epsilon \ll 1$ the result is a set of dynamical 
equations governing the collective coordinates $x^{(i)}$, $y^{(i)}$, 
and $\phi^{(i)}$, being the center positions along the $x$ and $y$-axis 
and accumulated phase of soliton $i$=1,2, respectively. 
We can express the new coordinates as
$x^{(i)}(z)=\int_0^z\nu_x^{(i)}(Z')\mbox{d}Z'+x_0^{(i)}$,
$y^{(i)}(z)=\int_0^z\nu_y^{(i)}(Z')\mbox{d}Z'+y_0^{(i)}$, and
$\phi^{(i)}(z)=\int_0^z\lambda^{(i)}(Z')\mbox{d}Z'+\phi^{(i)}_0$,
where subscript 0 denotes initial values.

At this point one traditionally simplifies the system by assuming the 
velocities to be small
$(\partial_z x^{(i)}\sim\epsilon,\;\partial_z y^{(i)}\sim\epsilon)$,
i.e. the solitons propagate almost in parallel. 
However, we are interested in velocities that can be considerable, so 
instead we assume symmetric interaction between in-phase solitons with 
initially identical profiles, $\lambda=\lambda^{(i)}$ and $P=P^{(i)}$, 
and equal but opposite velocities, $\nu_{x,y}=\nu_{x,y}^{(1)}=
-\nu_{x,y}^{(2)}$. 
Without loss of generality we set $x_0=x_0^{(1)}=-x_0^{(2)}\geq0$ and
$y_0=y_0^{(1)}=-y_0^{(2)}\geq0$. 
Symmetry is conserved and the two sets of collective coordinates degenerate 
to one, $X=x^{(1)}=-x^{(2)},Y=y^{(1)}=-y^{(2)}$. In cylindrical coordinates 
with $R=\sqrt{X^2+Y^2}$ we can then reduce the dynamical equations to the 
Euler-Lagrange equation of the effective Lagrangian 
\begin{eqnarray}
  L(R,\dot{R})=\frac{1}{2}P\dot{R}^2-U_{\mbox{{\small eff}}}(R,\dot{R}),
  \label{eqn:efflag}
\end{eqnarray}
for the single coordinate $R$. The effective potential 
\begin{eqnarray}
&&U_{\mbox{{\small
eff}}}(R,\dot{R})=\frac{C_0}{2R^2}P+\frac{1}{2}U(R,\dot{R})\label{eqn:effpot}
\end{eqnarray}
is composed of the classical centrifugal barrier, where
$C_0=(X\dot{Y}-Y\dot{X})^2=(x_0\nu_y+y_0\nu_x)^2$ is constant
because of conservation of angular momentum, and of the interaction
integral
\begin{eqnarray}
  && U=-\int V^{(1)}\bigg[V^{(1)}W^{(2)}\cos(2\phi)
     +2W^{(1)}V^{(2)}\cos(\phi)\bigg]d\vec{r}_\perp\,\nonumber\\
  \label{eqn:genpot}
\end{eqnarray}
where $\phi=2\nu_xx+2\nu_yy$. We note that, strictly speaking, $U$ is 
only a quasi-classical potential since it depends on the velocities 
(in contrast to the potential used in \cite{Ste98Bur99}).

We have now established a picture of an effective particle moving in 
a potential, $U_{\mbox{{\small eff}}}$, with the kinetic energy 
$E_{\mbox{\small kin}}=\frac{1}{2}P\dot{R}^2\geq0$. For small
velocities the potential (\ref{eqn:genpot}) has the shape of
a well and hence represents an attractive force. 
In the general case of non-planar interaction, $C_0\neq0$, the
 centrifugal barrier is always repulsive and goes to infinity at 
$R=0$. This does not necessarily rule out fusion
since also the velocities go to infinity because of conservation of
$C_0$. The centrifugal barrier also creates a local minimum in the
effective potential (still assuming small velocities) which suggests
that spiraling configurations may exist. In general, however,
we cannot expect our model to yield correct physical results in the vicinity of
$R=0$ since it violates the assumption of weakly overlapping
solitons. In fact fusion has been observed numerically, but stable
spiraling configurations have not been found\cite{Ste98Bur99}. 

Here we shall not discuss the qualitatively different regimes. Rather we
are interested in \emph{quantitative} predictions of escape velocities. 
For solitons launched with outwards velocities we will all ways be able 
to theoretically predict the escape
velocity. On the other hand, a consistent theory for the
determination of the inwards escape velocity only exists for in-plane
interaction, when the classical centrifugal barrier vanishes, i.e. $C_0=0$.

\begin{figure}
  \centerline{\hbox{
  \psfig{figure=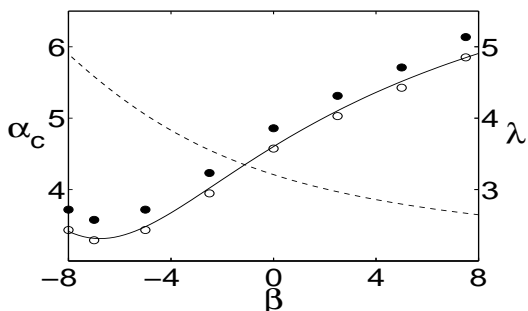,width=7cm,height=4cm}}\vspace{0.3cm}}
  \caption{Outwards escape angle in degrees (solid curve) for $FWHM$=1 and 
  $x_0$=1.5. Numerical experiments where exact solitons fused ($\bullet$) and 
  where they escaped ($\circ$). The dashed line shows the initial
  soliton power, $P$.}
  \label{fig:outwards_fig}
\end{figure}

We first determine the outwards escape angle. For simplicity we focus 
on in-plane interaction with $y_0=\nu_y=C_0=0$ wherefore $R=|X|$ and
$\dot{R}=\nu_x$. In this case the effective particle either escapes the
potential, $E_{\mbox{\small tot}}=E_{\mbox{\small kin}}+E_{\mbox{\small 
pot}}>0$, or is trapped by it, $E_{\mbox{\small tot}}<0$, and the escape 
velocity, $\nu_c$, is given by the relation 
\begin{equation}\label{eqn:transenergy}
  \nu_c^2P=U(x_0,\nu_c).
\end{equation}
Unfortunately we are not able to express the interaction integral
$U(x_0,\nu_x)$ in terms of analytical functions and we cannot
use Gaussians, since the Gaussian tale 
asymptotic is different from that of the exact soliton. 
It is however trivial to solve (\ref{eqn:transenergy}) numerically and
in Fig. \ref{fig:outwards_fig} we have plotted the outwards escape angle, 
given by $\alpha_c=\mbox{Arctan}(\nu_c)$, versus the phase mismatch.
The initial beam width and separation are kept constant to ensure a weak 
overlap of the soliton tails at all phase mismatches. 
The simulations were performed with numerically found
exact solitons as initial conditions. They confirm the accuracy of
the escape angle predicted by Eq.~(\ref{eqn:transenergy}). We found
the minimum of about 3$^\circ$ around $\beta=-7$ to be global. 
Note that the angles are expected to be small, since the initial 
overlap and hence the attractive force between the solitons is weak. 
As an example of the dynamics we show 
in Fig.~\ref{fig:gauss_beta0}:C,D the outcome of the experiments 
with $\beta$=5 from Fig.~\ref{fig:outwards_fig}. 
Only the fundamental waves are shown, the evolution of the SH waves 
being qualitatively the same.

Now considering solitons launched towards each other we first 
elaborate on the effective particle picture. If we assume the
solitons to be initially far apart this corresponds to the
effective particle experiencing essentially no potential. Even a small
launch angle should then result in a positive total energy and enable 
the effective particle to cross the bottom of the potential and escape 
towards infinity, corresponding to soliton crossing. This is
off course not the correct physical picture, since our system is not 
integrable and thus in reality the collision is not elastic. 
There is transfer of energy into internal soliton modes and shedding 
of energy as radiation. 

In a different picture we assume that the soliton profiles do not change
before the point of collision. This seems reasonable when comparing the 
characteristic length of slow adiabatic change with the relatively short 
interaction distance occurring for considerable velocities. 
In this case we can treat the interaction as if the solitons were launched 
on top of each other ($x_0$=0) corresponding to the effective particle 
being launched in the bottom of the potential, where it experiences the 
maximum barrier. 
Then the relation determining the escape velocity becomes 
$\nu_c^2P=2U(0,\nu_c)$ rather than (\ref{eqn:transenergy}). 
The interaction integral no longer
depends on the asymptotic tales but on the entire profiles and 
hence we can apply the Gaussian approximation (\ref{eqn:gaussapp}). 
The general transcendental equation for the inwards escape angle is 
then given by
\begin{eqnarray}
&&\nu_c^2=\frac{2}{b}\frac{\lambda b+1}{2\lambda b -1}
\bigg[
e^{-
{\textstyle \frac{4}{3}}b\nu_c^2}+2e^{-
{\textstyle \frac{1}{3}}b\nu_c^2}\bigg],\label{eqn:transenergygen}
\end{eqnarray}
which for $\beta$=0 simplifies to 
\begin{equation}
  \beta=0 : \quad \nu_c = 0.23\times\sqrt{P},
  \label{eqn:transinw}
\end{equation}
in terms of the power. In the large phase-mismatch cascading limit, 
($\beta\gg\lambda$) where the nonlinearity is effectively cubic, 
Eq.~(\ref{eqn:transenergygen}) simplifies to
\begin{equation}
  \beta\gg\lambda : \quad 
  \nu_c = \sqrt{\frac{3}{4}\left(\frac{P}{2\pi}-\beta\right)}.
  \label{eqn:transinwcasc}
\end{equation}
We remark that this approach is equivalent to finding the critical
angle of total internal reflection for a waveguide \cite{Sny91}. 
However, since beam propagation in quadratic media does not induce 
changes in the refractive index the method used in \cite{Sny91} 
is not applicable to this case. 

\begin{figure}
  \centerline{\hbox{
  \psfig{figure=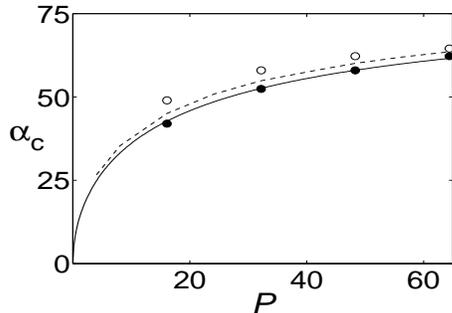,width=6cm,height=4cm}}\vspace{0.2cm}}
  \caption{Inwards escape angle in degrees versus soliton power calculated
  analytically with Gaussians (solid) and with exact solitons (dashed) for 
  $\beta$=0 and $x_0$=5. Numerical experiments where Gaussians crossed 
  ($\circ$) and fused ($\bullet$).}
  \label{fig:inwards_beta0}
\end{figure}

\begin{figure}
  \centerline{\hbox{
  \psfig{figure=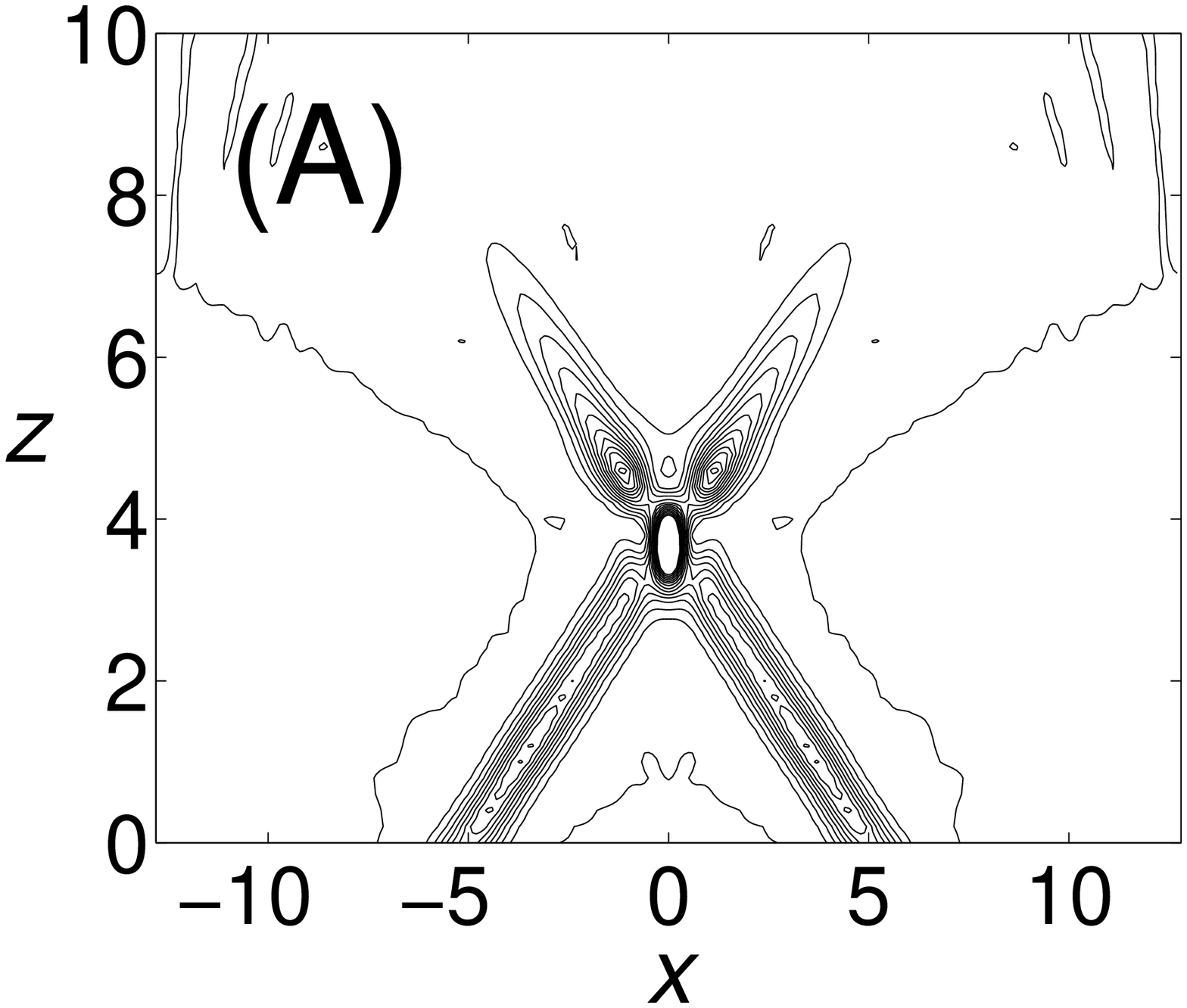,width=3.5cm,height=3.0cm}
  \psfig{figure=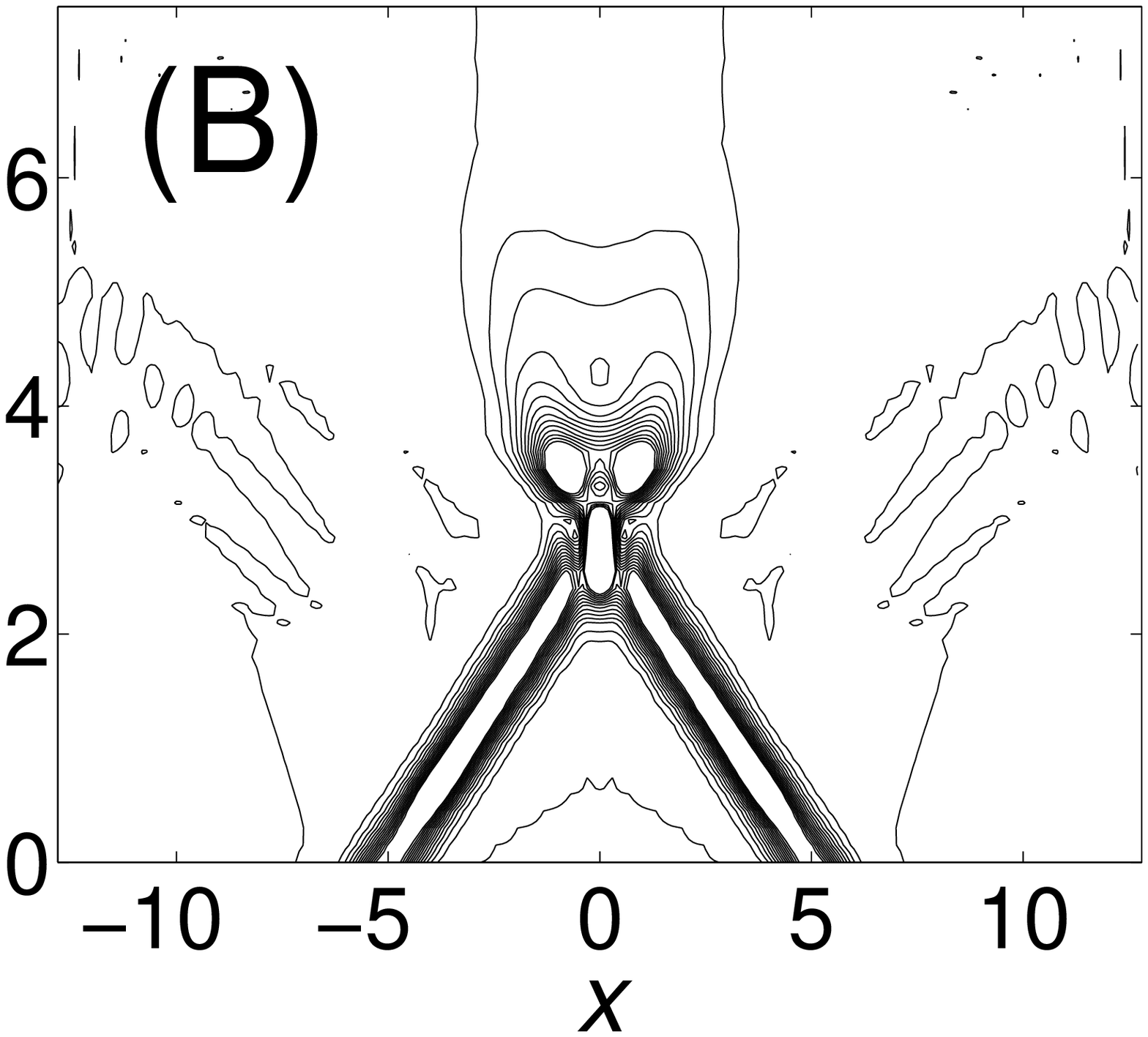,width=3.5cm,height=3.0cm}}\vspace{0.2cm}}
  \caption{A: Crossing followed by diffraction ($\beta=-3$ and $P=45.3$). 
  B: Annihilation ($\beta=0$ and $P=48.3$).}
  \label{fig:runs_ex}
\end{figure}

In Fig.~\ref{fig:inwards_beta0} we have summarized the results for 
exact phase matching, $\beta$=0, and plotted the predicted inwards 
escape angle, $\alpha_c=\mbox{Arctan}(\nu_c)$, versus soliton
power, both for $\nu_c$ given by (\ref{eqn:transinw}) and for $\nu_c$ 
found with exact solitons. The curves are close and the simple square root dependency on the power excellently predict the escape angle. In the
experiments we used Gaussians as initial conditions. These were launched with a distance of $2x_0$=10
between them, ensuring practically zero initial overlap. In
Fig. \ref{fig:gauss_beta0}:A,B we show examples of experiments with
$\beta=0$ and a power of $P=48.3$. We note that
for the exact soliton initial conditions we observed even better
agreement than with Gaussians. Close to the escape
angle all the power is shed as radiation and thus (\ref{eqn:transinw})
serves as an
\emph{accurate prediction of soliton annihilation} (Fig.~\ref{fig:runs_ex}:B). 
We also investigated the cases of non-zero mismatches, focusing on 
$\beta=\pm3$. In these regimes there is a power threshold for soliton 
excitation and the collisions are of a much more complex nature than 
for perfect phase matching, where solitons exist at all powers. 
For relatively low powers not far above the threshold the collisions 
mostly resulted in destruction of the solitons (Fig.~\ref{fig:runs_ex}:A). 
The explanation of this phenomenon is that too much power is shed as 
radiation in the collision and hence the resulting beams diffract 
because they do not carry sufficient power to form solitons. 
For higher powers the predicted escape angles were reasonably close 
to the observations.

In conclusion we have developed a theoretical description that 
should hold for systems with all types of local nonlinearities. 
In particular 
we have studied bulk $\chi^{(2)}$ media and determined analytical 
expressions for the escape angles when the centrifugal barrier
vanishes. This happens in the two in-plane cases of outwards and
inwards launched solitons.  
The simple expression for the inwards escape angle represents the
first {\em analytical} prediction of the {\em outcome} of a soliton 
collision.  We have verified the analytical expressions
numerically using Gaussian approximations and observed 
excellent agreement.

We acknowledge support from the Danish Technical Research Council
under Talent Grant No. 56-00-0355. Much of the numerical work was
carried out at Centre de Computacio i Comunicacions de Catalunya
\bibliographystyle{prsty}

\end{multicols}

\end{document}